\documentclass[twocolumn,showpacs,amssymb,amsmath,aps,pra]{revtex4}
\newcommand{\half}{\mbox{$\frac{1}{2}$}}
\usepackage{graphics}
\usepackage{bm}
\usepackage{graphics}
\usepackage{bm}
\usepackage{amsmath}
\usepackage{amssymb}
\begin{document}

\title{Factorization and entanglement in general $XYZ$ spin arrays in
non-uniform transverse fields}
\author{R. Rossignoli, N. Canosa, J.M. Matera}
\affiliation{Departamento de F\'{\i}sica-IFLP,
Universidad Nacional de La Plata, C.C. 67, La Plata (1900) Argentina}
\date{\today}
\begin{abstract}

We determine the conditions for the existence of a pair of degenerate parity
breaking separable eigenstates in general arrays of arbitrary spins connected
through $XYZ$ couplings of arbitrary range and placed in a transverse field,
not necessarily uniform. Sufficient conditions under which they are ground
states are also provided. It is then shown that in finite chains, the
associated definite parity states, which represent the actual ground state in
the immediate vicinity of separability, can exhibit entanglement between any
two spins regardless of the coupling range or separation, with the reduced
state of any two subsystems equivalent to that of pair of qubits in an
entangled mixed state. The corresponding concurrences and negativities are
exactly determined. The same properties persist in the mixture of both definite
parity states. These effects become specially relevant in systems close to the
$XXZ$ limit. The possibility of field induced alternating separable solutions
with controllable entanglement side limits is also discussed. Illustrative
numerical results for the negativity between the first and the $j^{\rm th}$
spin in an open spin $s$ chain for different values of $s$ and $j$ are as
well provided.
\end{abstract}
\pacs{03.67.Mn, 03.65.Ud, 75.10.Jm}
\maketitle

Quantum entanglement constitutes one of the most fundamental, complex and
counter-intuitive aspects of quantum mechanics. It is an essential resource in
quantum information theory \cite{NC.00}, playing a key role in quantum
teleportation \cite{Be.93} and computation \cite{NC.00,BD.00,RB.01}. It also
provides a deeper understanding of quantum correlations in many-body systems
\cite{AFOV.08}. In particular, a great effort has been devoted in recent years
to analyze entanglement and its connection with critical phenomena in spin
chains \cite{AFOV.08,ON.02,OO.02,VV.03}. Studies of {\it finite} chains, of
most interest for quantum information applications, are presently also
motivated by the possibility of their controllable simulation through quantum
devices \cite{HBP.07,CAB.08}.

A remarkable feature of interacting spin chains is the possibility of
exhibiting {\it exactly separable} ground states (GS) for special values of the
external magnetic field, first discovered in  \cite{KTM.82,MS.85} in a $1D$
$XYZ$ chain with first neighbor coupling. It was recently investigated in more
general arrays under uniform fields
\cite{RR.04,DV.05,AA.06,GI.07,RCM.08,GAI.08,GG.09}, with a completely general
method for determining separability introduced in \cite{GAI.08}. Another
remarkable related aspect is the fact that in the immediate vicinity of these
separability points (SP) the entanglement between two spins can reach {\it
infinite range} \cite{AA.06,RCM.08}. In \cite{RCM.08} we have shown that the SP
in finite cyclic spin $1/2$ arrays in a transverse field corresponds actually
to a GS transition between opposite parity states (the last level crossing for
increasing field), with the entanglement between {\it any} two spins reaching
there finite side limits irrespective of the coupling range. In a small chain,
this SP plays then the role of a ``quantum critical point''. In contrast, the
entanglement range remains typically finite and low at the conventional phase
transition \cite{ON.02}.

The aim of this work is to generalize previous results to $XYZ$ arrays of {\it
arbitrary} spins and geometry in a {\it general} transverse field, not
necessarily uniform.  Moreover, we will also determine the exact side limits of
the entanglement between {\it any} two subsystems (including those for the
block entropy and those for any two spins or group of spins) at the SP
analytically, for {\it any} spin value. A non-uniform field will be shown to
allow exact separability with infinite entanglement range in its vicinity in
quite diverse systems (such as open or non-uniform chains), including the
possibility of {\it field induced alternating separable solutions} along
separability curves, with {\it controllable} entanglement side limits.
Illustrative results for the negativity between the first and $j^{\rm th}$ spin
in an open spin $s$ chain as a function of field and separation are as well
presented, for different spin values.

We consider $n$ spins $\bm{s}_i$ (which can be regarded as qudits of dimension
$d_i=2s_i+1\geq 2$) not necessarily equal, interacting through $XYZ$ couplings
of arbitrary range in the presence of a transverse external field $b^i$, not
necessarily uniform. The Hamiltonian reads
\begin{equation}
H=\sum_i b^i s^z_i-\half\sum_{i,j}(v_x^{ij} s_i^xs_j^x+
v_y^{ij}s_i^ys_j^y+v_z^{ij}s_i^zs_j^z)\,,\label{H1}
\end{equation}
and commutes with the global $S_z$ parity or phase-flip
$P_z=\exp[i\pi\sum_{i=1}^n (s_i^z+s_i)]$ for any values of $b^i$, $v_\mu^{ij}$
or $s_i$. Self-energy terms ($i=j$), non-trivial for $s_i\geq 1$, are for
instance present in recent coupled cavity based simulations of arbitrary spin
$XXZ$ models \cite{CAB.08} and will be allowed if $s_i\geq 1$.

We now seek the conditions for which such system will possess a {\it separable
parity breaking eigenstate} of the form
\begin{eqnarray}
|\Theta\rangle&=&\otimes_{i=1}^n
\exp[i\theta_i s^y_i]|0_i\rangle\label{Th}\\
&=&\otimes_{i=1}^n[\sum_{k=0}^{2s_i}\sqrt{(^{2s_i}_{\,k})}
\cos^{2s_i-k}{\textstyle\frac{\theta_i}{2}}
\sin^{k}{\textstyle\frac{\theta_i}{2}}|k_i\rangle]\,,\label{Thex}
\end{eqnarray}
where $s^z_i|k_i\rangle=(k-s_i)|k_i\rangle$ and $e^{i\theta_i
s^y_i}|0_i\rangle$ is a rotated minimum spin state (coherent state
\cite{ACGT.72}). The choice of $y$ as rotation axis does not pose a loss of
generality as any state $e^{i\bm{\phi}_i\cdot\bm{s}_i}|0_i\rangle$ corresponds
to a suitable complex $\theta_i$ in (\ref{Th}) \cite{r1}. Replacing $s_i^\mu$
in (\ref{H1}) by $e^{-i\theta_i s_i^y}s_i^\mu e^{i\theta_is_i^y}$, i.e.,
$s_i^{z,x}\rightarrow s_i^{z,x}\cos\theta_i\pm s_i^{x,z}\sin\theta_i$,
$s_i^y\rightarrow s_i^y$, the equation
$H|\Theta\rangle=E_\Theta|\Theta\rangle$, i.e., $H_\Theta|0\rangle=E_\Theta
|0\rangle$ with $|0\rangle=\otimes_{i=1}^n|0_i\rangle$ and
$H_\Theta=e^{-i\sum_i\theta_i s_i^y}He^{i\sum_i\theta_i s_i^y}$, leads to the
equations
\begin{eqnarray}
v_y^{ij}&=&v_x^{ij}\cos\theta_i\cos\theta_j+v_z^{ij}
\sin\theta_i\sin\theta_j\,,
\label{a}\\b^i\sin\theta_i&=&\sum_{j}
(s_j-\half\delta_{ij})(v_x^{ij}\cos\theta_i\sin\theta_j
-v_z^{ij}\sin\theta_i\cos\theta_j) \,,\label{b}
\end{eqnarray}
which determine, for instance, the values of $v_y^{ij}$ and $b^i$ in terms of
$v_x^{ij}$, $v_z^{ij}$, $s_i$ and $\theta_i$. The energy is then given by
\begin{eqnarray}
E_\Theta&=&-\sum_i s_i [b^i\cos\theta_i+
\half\sum_{j}(s_j-\half\delta_{ij})(v_x^{ij}\sin\theta_i\sin\theta_j
\nonumber\\
&&+v_z^{ij}\cos\theta_i\cos\theta_j)+
{\textstyle\frac{1}{4}}(v^{ii}_x+v^{ii}_y+v^{ii}_z)]\,.\label{en}
\end{eqnarray}
For a $1D$ spin $s$ cyclic chain with first neighbor couplings
($v_\mu^{ij}=v_\mu \delta_{i,j\pm 1}$) in a uniform field ($b^i=b$) we recover
the original GS separability conditions of ref.\ \cite{MS.85} for both the
ferromagnetic ($v_\mu\geq 0$, $\theta_i=\theta$) and antiferromagnetic
($v_\mu\leq 0$, $\theta_i=(-1)^i\theta$) cases. Eqs.\ (\ref{a})--(\ref{en}) are
however completely general and actually hold also for  {\it complex} values of
$\theta_i$, $v_\mu^{ij}$ and $b^i$: If satisfied $\forall i,j$, $H$ will have a
separable eigenstate (\ref{Th}) with eigenvalue (\ref{en}). If
$\sin\theta_i\neq 0$ for some $i$, this eigenvalue is {\it degenerate}:
$|\Theta\rangle$ will break parity symmetry and therefore, the partner state
\begin{equation}
|-\Theta\rangle=P_z|\Theta\rangle=\otimes_{i=1}^n \exp[-i\theta_is^y_i]
|0_i\rangle\,,\label{Thr}
\end{equation}
will be an exact eigenstate of $H$ as well, with the same energy (\ref{en}).
The points in parameter space where the states $|\pm\Theta\rangle$ become exact
eigenstates correspond necessarily to the crossing of at least two opposite
parity levels.

For real $\theta_i$, Eq.\ (\ref{b}) is just the stationary condition for the
energy (\ref{en}) at fixed $b^i$, $v_\mu^{ij}$. The state (\ref{Th}) can thus
be regarded as a mean field trial state, with Eq.\ (\ref{b}) the associated
self-consistent equation. Eq.\ (\ref{a}), which is spin independent (at fixed
$v_\mu^{ij}$), ensures that it becomes an exact eigenstate by canceling the
residual one and two-site matrix elements connecting $|\Theta\rangle$ with the
remaining states. Moreover, if $\theta_i\in(0,\pi) \forall i$ and
\begin{equation}
|v_y^{ij}|\leq v_x^{ij}\;\;\;\forall\,i,j\,,\label{ve0}
\end{equation}
we can ensure that $|\!\!\pm\Theta\rangle$ will be {\it ground states} of $H$:
In the standard basis formed by the states $\{\otimes_{i=1}^n|k_i\rangle\}$,
the terms in $H$ depending on $\{s^z_i\}$ are diagonal whereas the rest lead to
real non-positive off-diagonal matrix elements, as
$\sum_{\mu=x,y}\!\!v_\mu^{ij}s^\mu_{i}s^\mu_j=
\sum_{\nu=\pm}\!v_\nu^{ij}(s^+_is^{-\nu}_j +s^-_is^{\nu}_j),$ where
$s_j^\pm=s_j^x\pm is_j^y$ and $v_\pm^{ij}=\frac{1}{4}(v_x^{ij}\pm v_y^{ij})\geq
0$ by Eq.\ (\ref{ve0}). Hence, $\langle H\rangle$ can be minimized by a state
with all coefficients real and of the same sign in this basis (different signs
will not decrease $\langle H\rangle$), which then, cannot be orthogonal to
$|\Theta\rangle$ (Eq.\ (\ref{Thex})). With suitable phases for $\theta_i$,
$|\pm\Theta\rangle$ can also be GS in other cases: A $\pi$ rotation around the
$z$ axis at site $i$ leads to $\theta_i\rightarrow -\theta_i$ and
$v_{x,y}^{ij}\rightarrow -v_{x,y}^{ij}$ for $i\neq j$.

{\it Definite parity eigenstates of $H$} in the subspace generated by the
states $|\pm\Theta\rangle$ can be constructed as
\begin{eqnarray}
|\Theta^\pm\rangle&=&\frac{|\Theta\rangle\pm|-\Theta\rangle} {\sqrt{2(1\pm
O_\Theta)}}\label{Ps}, \\
O_\Theta&\equiv&\langle-\Theta|\Theta\rangle=
{\textstyle\prod_{i=1}^n}\cos^{2s_i}\theta_i\,, \label{ov}
\end{eqnarray}
which satisfy $P_z|\Theta^{\pm}\rangle=\pm |\Theta^{\pm}\rangle$,
$\langle\Theta^\nu|\Theta^{\nu'}\rangle=\delta^{\nu\nu'}$. Here we have set
$\theta_i$ real $\forall i$, since by local rotations around the $z$ axis we
can always choose $y_i$ in the direction of ${\bm{\phi}}_{i}$ (and hence
$\theta_i$ real) in the final state $|\Theta\rangle$. Moreover, we may also set
$|\theta_i|\leq \pi/2$ (and hence $O_\Theta\geq 0$) since a local rotation of
$\pi$ around the $x$ axis leads to $\theta_i\rightarrow\pi-\theta_i$. The
overlap (\ref{ov}) will play an important role in the following.

When the degeneracy at the SP is indeed 2, the states (\ref{Ps}) (rather than
(\ref{Th})) are {\it the actual side limits at the SP} of the corresponding
non-degenerate (and hence definite parity) exact eigenstates of $H$. For small
variations $\delta b^i$, the degeneracy will be broken if $O_\Theta\neq 0$,
with an energy gap given by $\Delta E\approx\sum_i\delta b^i \Delta M_i$, where
\[\Delta M_i\equiv\langle\Theta^-|s_i^z|\Theta^-\rangle-
\langle \Theta^+|s_i^z|\Theta^+\rangle= \frac{2s_i\sin^2\theta_i\,
 O_\Theta}{\cos\theta_i(1-O^2_\Theta)}\,.\]
(In contrast, $\langle\pm\Theta|s_i^z|\!\pm\Theta\rangle=-s_i\cos\theta_i$).
When $|\Theta^\pm\rangle$ are GS, a GS parity transition
$|\Theta^-\rangle\rightarrow|\Theta^+\rangle$, characterized by a {\it
magnetization step} $\Delta M=\sum_i \Delta M_i$, will then take place at the
SP if all or some of the fields are increased across the factorizing values
(\ref{b}). If $\Delta E$ or $\Delta M$ can be resolved or measured, the
realization of the states (\ref{Ps}) is then ensured. Their magnitude is
governed by the overlap (\ref{ov}), appreciable in small systems (if
$\theta_i\neq \pi/2$) as well as in finite systems with small angles
$\theta_i^2\approx \delta_i/n$, such that $O_\Theta\approx e^{-\sum_i
s_i\delta_i/n}$. This implies (Eq.\ (\ref{a})) systems close to the $XXZ$ limit
($v_y^{ij}=v_x^{ij}$). In this limit ($\theta_i\rightarrow 0$), $\Delta
M\rightarrow 1$, with $|\Theta^+\rangle\rightarrow |0\rangle$ and
$|\Theta^-\rangle\propto\sum_i \sqrt{s_i}\theta_i|1_i\rangle$ (weighted
$W$-type state), where
$|1_i\rangle\equiv\otimes_{j=1}^n|(\delta_{ji})_j\rangle$.

{\it Entanglement of definite parity states}.  In contrast with
$|\!\!\pm\Theta\rangle$, the states (\ref{Ps}) are entangled. If
$\sin\theta_i\neq 0\,\forall$ $i$ the Schmidt number for {\it any} global
bipartition $(A,\bar{A})$ is $2$ and the Schmidt decomposition is
\begin{eqnarray}
|\Theta^{\pm}\rangle&=&
\sqrt{p^{\pm}_{A^+}}|\Theta^+_A\rangle|\Theta^{\pm}_{\bar{A}}\rangle
+\sqrt{p^\pm_{A^-}}|\Theta^{-}_A\rangle|\Theta^{\mp}_{\bar{A}}\rangle\,,
\label{s1} \\
p^\pm_{A^\nu}&=&{\textstyle\frac{(1+\nu O_{A})(1\pm\nu O_{\bar{A}})}{2(1\pm
O_\Theta)}}\,, \;\;O_A=\langle-\Theta_A|\Theta_A\rangle\,,
\end{eqnarray}
where $|\Theta^\pm_{A}\rangle$, $|\Theta^\pm_{\bar{A}}\rangle$ denote the
analogous normalized definite parity states for the subsystems $A$, $\bar{A}$,
with $\nu=\pm$, $O_AO_{\bar{A}}=O_\Theta$ and $p^\pm_{A^+}+p^{\pm}_{A^-}=1$.
Hence, $|\Theta^\pm\rangle$ can be effectively considered as {\it two qubit
states} with respect to {\it any} bipartition $(A,\bar{A})$, with
$|\Theta_A^\pm\rangle$, $|\Theta_{\bar{A}}^\pm\rangle$ representing the
orthogonal states of each qubit. Accordingly, the reduced density matrix
$\rho_A^\pm$ of subsystem $A$ in the state $|\Theta^\pm\rangle$ is
\begin{equation}
\rho^\pm_A=p^\pm_{A^+}|\Theta^+_A\rangle\langle
\Theta^+_A|+p^\pm_{A^-}|\Theta^-_A\rangle\langle\Theta^-_A|
 \label{rhoa}\,.\end{equation}
The entanglement between $A$ and its complement $\bar{A}$ can be measured
through the global concurrence (square root of the tangle \cite{CC.03})
$C_{A\bar{A}}=\sqrt{2(1-{\rm tr}\,\rho^2_A)}$, which for a rank 2 density is
just an increasing function of the entanglement entropy $E_{A\bar{A}}=-{\rm
tr}\,\rho_A\log_2\rho_A$, with $C_{A\bar{A}}=E_{A\bar{A}}=0$ ($1$) for a
separable (Bell) state. In the states (\ref{s1}) we then obtain
\begin{equation}
C^\pm_{A\bar{A}}=\frac{\sqrt{(1-O_A^2)(1-O_{\bar{A}}^2)}}
 {1\pm O_\Theta} \label{Ca}\,.
  \end{equation}
These values represent the side limits of $C_{A\bar{A}}$ at the SP. For
$O_\Theta>0$, $C_{A\bar{A}}^->C_{A\bar{A}}^+$, with $C_{A\bar{A}}^-=1$ if
$O_A=O_{\bar{A}}$. Note  that $|\Theta^\pm\rangle$ are simultaneous Bell states
for $(A,\bar{A})$ only if $O_A=O_{\bar{A}}=0$ (GHZ limit of
$|\Theta^\pm\rangle$). Increasing overlaps will in general decrease the global
entanglement.

At the SP, the entanglement entropy of a block of $L$ spins in a $1D$ first
neighbor spin $1/2$ $XY$ chain in a constant field was found in \cite{KK.07} to
be $S_L=-{\rm tr}\rho_L\ln\rho_L=\ln 2$ (i.e., $C_{L\bar{L}}=E_{L\bar{L}}=1$)
in the thermodynamic limit, in agreement with Eq.\ (\ref{Ca}) for vanishing
overlaps. Eq.\ (\ref{Ca}) extends this result to general {\it finite} chains,
leading to a slightly smaller value: For small $O_A$, $O_{\bar{A}}$,
$C^\pm_{A\bar{A}}\approx 1-\half (O_A\pm O_{\bar{A}})^2$ and $S^\pm_{L}\approx
\ln 2-\half (O_L\pm O_{\bar{L}})^2$ (with $O_L=(\frac{v_y}{v_x})^{\frac{L}{2}}$
in the $s=1/2$ $XY$ chain).

{\it Pairwise and subsystem entanglement}. On the other hand, the entanglement
of a subsystem is enabled by non-zero overlaps. A remarkable feature of the
states (\ref{Ps}) is that {\it any} two spins or disjoint subsystems $B,C$ {\it
will also be entangled} if the complementary overlap $O_{\overline{B+C}}$ is
non-zero and $O_B^2<1$, $O_C^2<1$. Moreover, this entanglement can be
characterized by the concurrence
\begin{equation}
C^\pm_{BC}=\frac{\sqrt{(1-O_B^2)(1-O_C^2)}O_{\overline{B+C}}} {1\pm
O_\Theta}\,,\label{Cbc}
\end{equation}
or equivalently, the negativity \cite{VW.02,ZHSL.98},
\begin{equation}
N_{BC}^\pm=\half[\sqrt{(p^{\pm}_{A^{\mp}})^2
+(C_{BC}^{\pm})^2/O_{\overline{B+C}}}-p^{\pm}_{A^\mp}]\,, \label{Neg}
\end{equation}
where $A=B+C$. While the concurrence of an arbitrary mixed state $\rho_{A}$
(which can be defined through the convex roof extension of the pure state
definition \cite{DC.07}) is not directly computable in general (the exception
being the case of two qubits \cite{HW.97}), the negativity $N_{BC}= \half[{\rm
Tr}|\rho_{A}^{t_B}|-1]$, where $\rho_A^{t_B}$ denotes partial transpose
\cite{PP.93}, can always be calculated \cite{RC.06}, being just the absolute
value of the sum of the negative eigenvalues of $\rho_A^{t_B}$. Eq.\
(\ref{Neg}) represents then the side-limits of $N_{BC}$ at the SP.

{\it Proof:} For $A=B+C$, we first note that if $O_{\bar{A}}=0$, Eq.\
(\ref{rhoa}) becomes $\rho_A^\pm=\half(|\Theta_A\rangle\langle
\Theta_A|+|\!-\Theta_A\rangle\langle-\Theta_A|)$, i.e., $\rho_A^\pm$ {\it
coincident and} {\it separable} (convex combination of product densities
\cite{WW.89}). Entanglement between $B$ and $C$ can then only arise if
$O_{\overline{B+C}}\neq 0$. Next, using similar Schmidt decompositions
(\ref{s1}) of the states $|\Theta_A^\pm\rangle$, Eq.\ (\ref{rhoa}) can also be
considered as an {\it effective two-qubit mixed state} with respect to {\it
any} bipartition $(B,C)$ of $A$: Its support will lie in the subspace spanned
by the four states
$\{|\Theta_B^\nu\rangle|\Theta_C^{\nu'}\rangle\,,\nu,\nu'=\pm\}$, such that
\[\rho_A^\pm=\left(\begin{array}{cccc}
p^\pm_{A^+}q^+_{BC^+}&0&0&p^\pm_{A^+}\alpha^+_{BC}\\
0&p^\pm_{A^-}q^-_{BC^+}&p^\pm_{A^-}\alpha^-_{BC}&0\\
0&p^\pm_{A^-}\alpha^-_{BC}&p^\pm_{A^-}q^-_{BC^-}&0\\
p^\pm_{A^+}\alpha^+_{BC}&0&0&p^\pm_{A^+}q^+_{BC^-}
 \end{array}\right)\]
where $q^\pm_{BC^\nu}=\frac{(1+\nu O_B) (1\pm\nu O_C)} {2(1\pm O_BO_C)}$,
$\alpha^\pm_{BC}=\sqrt{q^\pm_{BC^+}q^\pm_{BC^-}}$ and
$q^\pm_{BC^+}+q^\pm_{BC^-}=1$. $\rho_A^\pm$ will be entangled if its partial
transpose has a negative eigenvalue \cite{PP.93}, a condition here equivalent
to a positive mixed state concurrence \cite{HW.97} $C^\pm_{BC}= {\rm
Max}[C^\pm_+,C^\pm_-,0]$, where $C^\pm_{\nu}=
2[p_{A^\nu}^\pm\alpha_{BC}^\nu-p_{A^{-\nu}}^\pm\alpha_{BC}^{-\nu}]$ represent
parallel ($\nu=+$) or antiparallel ($\nu=-$) concurrences, i.e. driven by
$|\Theta_A^+\rangle$ or $|\Theta_A^-\rangle$ in Eq.\ (\ref{rhoa}). This leads
to Eq.\ (\ref{Cbc}), with $C^+_{BC}$ ($C^-_{BC}$) {\it parallel} ({\it
antiparallel}). The ensuing negativity, given here by minus the negative
eigenvalue of the partial  transpose $(\rho_A^\pm)^{t_B}$, is then given by
Eq.\ (\ref{Neg}).

For $B=A$, $C=\bar{A}$ ($O_{\overline{B+C}}\rightarrow 1$), Eq.\ (\ref{Cbc})
reduces to (\ref{Ca}), with $N_{A\bar{A}}^\pm=\half C^{\pm}_{A\bar{A}}$. For a
pair of spins $i\neq j$, $O_B=\cos^{2s_i}\theta_i$, $O_C=\cos^{2s_j}\theta_j$
and the result of \cite{RCM.08} is recovered from (\ref{Cbc}) if $s_i=\half$
and $\theta_i=\theta$ $\forall i$. We finally note that if $O_B=O_C$,
$N^+_{BC}=C^+_{BC}/2$, as in the case of a global partition. In general,
however, there is no proportionality between $N_{BC}^{\pm}$ and $C_{BC}^\pm$.

The concurrences (\ref{Cbc}) fulfill the monogamy inequalities \cite{CKW.00}
$C_{B,C+D}^2\geq C_{BC}^2+C_{BD}^2$ for any three disjoint subsystems $B,C,D$.
We actually obtain here
\begin{equation}
{\textstyle C_{BC}^2+C_{BD}^2=
C_{B,C+D}^2[1-\frac{(1-O_C^2)(1-O_D^2)}{1-O_C^2O_D^2}]}\,.
\end{equation}

Let us also remark that  subsystem entanglement persists, though attenuated, in
the uniform mixture
\begin{equation}
\rho^0=\half(|\Theta^+\rangle\langle\Theta^+|+|\Theta^-\rangle\langle\Theta^-|)
 \label{rho0}\,,\end{equation}
which differs from $\half(|\Theta\rangle\langle\Theta|
+|\!\!-\!\Theta\rangle\langle\!-\Theta|)$ if $O_\Theta\neq 0$ and represents
the $T\rightarrow 0^+$ limit of the thermal state $\rho\propto e^{-\beta H}$ at
the SP when $|\pm\Theta\rangle$ are GS (and the GS degeneracy there is 2).
Replacing $p^\pm_{A^\nu}$ by $\half(p^+_{A^\nu}+p^-_{A^\nu})$ in (\ref{rhoa}),
we find now {\it antiparallel global and subsystem concurrences}, given for any
disjoint subsystems $B,C$ by
\begin{equation}
C_{BC}^0=\half(C^-_{BC}-C^+_{BC})=C^-_{BC}O_\Theta/(1+O_\Theta)
\,,\label{Cbc0}
\end{equation}
i.e., half the parity splitting of $C_{BC}$. Eq.\ (\ref{Cbc0}) remains valid
for a global bipartition ($B=A$, $C=\bar{A}$). The ensuing negativity can be
similarly calculated.

The order of magnitude of subsystem concurrences is governed by
the complementary overlap $O_{\overline{B+C}}$. For small subsystems (like a
pair of spins) in a large system, $C_{BC}^\pm$ will be  appreciable just for
sufficiently small angles in the complementary system, i.e.,
$\theta_{i}^2\approx \delta_i/n$, such that $O_{\overline{B+C}}\approx
e^{-\sum_{i\in\bar{A}}s_i\delta_i/n}$ remains finite. This leads again to
systems with small $XY$ anisotropy.

{\it Uniform Solution}. Let us now examine the possibility of a common angle
$\theta_i=\theta$ $\forall$ $i$. Eq.\ (\ref{a}) leads then to
\begin{equation}
v_y^{ij}-v_z^{ij}=(v_x^{ij}-v_z^{ij})\cos^2\theta\,,\label{chi1}
\end{equation}
implying a fixed ratio $\chi\equiv
(v_y^{ij}-v_z^{ij})/(v_x^{ij}-v_z^{ij})=\cos^2\theta$ for {\it all} pairs with
$v_x^{ij}\neq v_z^{ij}$, and an isotropic coupling $v_y^{ij}=v_x^{ij}$ if
$v_x^{ij}=v_z^{ij}$. A subset of isotropic couplings will not spoil this
eigenstate \cite{r2}. Eq.\ (\ref{b}) implies then $b^i$ arbitrary if $\theta=0$
or $\pi$ ($XXZ$ case $v_y^{ij}=v_x^{ij}$) or otherwise
\begin{equation}
b^i=\cos\theta\sum_{j}(v_x^{ij}-v_z^{ij})(s_j-\half\delta_{ij}).
 \label{bs}\end{equation}
 The energy (\ref{en}) becomes
\begin{equation}
E_\Theta=-\half\sum_{i,j}s_i[s_j(v_x^{ij}+v_y^{ij}-v_z^{ij})
+\delta_{ij}v_z^{ii}]\,. \label{en2}
\end{equation}
A general field allows then a uniform separable eigenstate (a global coherent
state) in cyclic as well as open chains with arbitrary spins $s_i$ in any
dimension if (\ref{chi1}) holds $\forall i,j$. For instance, in an open $1D$
spin $s$ chain with first neighbor couplings $v_\mu^{ij}=
v_\mu\delta_{i,j\pm 1}$, Eq.\ (\ref{bs}) yields
$b^i=b_s=2s\sqrt{(v_y-v_z)(v_x-v_z)}$ at inner sites but $b^1=b^n=\half
b_s$ {\it at the borders}.

Eqs.\ (\ref{chi1})--(\ref{en2}) are actually valid for general complex
$\theta$, but real fields imply $\cos\theta$ real ($\chi\geq 0$). The case
$\cos^2\theta>1$ (imaginary $\theta$) corresponds to a rotation around the $x$
axis but can be recast as a rotation around the $y$ axis by a global rotation
around the $z$ axis. Hence, we may set $\cos^2\theta\in[0,1]$.
$|\pm\Theta\rangle$ will then be GS when Eq.\ (\ref{ve0}) holds.

The concurrence (\ref{Cbc}) becomes, setting $\cos^2\theta=\chi$,
\begin{eqnarray}
C^{\pm}_{BC}&=&\frac{\sqrt{(1-\chi^{2S_B})(1-\chi^{2S_C})}\chi^{S-(S_B+S_C)}}
 {1\pm\chi^S}\label{cgbcu}\end{eqnarray}
where $S_B=\sum_{i\in B}s_i$ is the subsystem total spin and $S=\sum_{i}s_i$
the total spin. It is independent of separation and coupling range, depending
solely on $\chi^S$ and the ratios $S_B/S$, $S_C/S$. If $\chi=1-\delta/2S$, with
$\delta>0$ and finite, $\chi^S\approx e^{-\delta/2}$ remains finite for large
$S$. Eq.\ (\ref{cgbcu}) leads then to $O(1/\sqrt{S})$ and $O(1/S)$ global and
subsystems concurrences for small $S_A$, $S_B$ and $S_C$:
\begin{eqnarray}
 C^\pm_{A\bar{A}}&\approx&\sqrt{\frac{S_A\delta}{S}}\frac{
\sqrt{1-e^{-\delta}}}{1\pm e^{-\delta/2}}\,,\\
C^\pm_{BC}&\approx&\frac{\delta}{S}\frac{\sqrt{S_B S_C} e^{-\delta/2}}{1\pm
 e^{-\delta/2}}\,.\label{cgbcu2}\end{eqnarray}

On the other hand, for $S_A=\half S$, $C^-_{A\bar{A}}=1$ whereas
$C^+_{A\bar{A}}=\tanh\frac{1}{4}\delta$. Thus, while for large $\delta$ both
$C^\pm_{A\bar{A}}$ rapidly approach 1 as $S_A$ increases, for small $\delta$
($XXZ$ limit) this occurs just for $C^-_{A\bar{A}}$ and $S_A$ close to $S/2$
(here $|\Theta^+\rangle\rightarrow |0\rangle$ but $|\Theta^-\rangle$ approaches
the $W$-type state $\propto\sum_i \sqrt{s}_i|1_i\rangle$).

{\it Alternating solution and controllable entanglement at the SP}. Among other
possibilities allowed by Eqs.\ (\ref{a})--(\ref{b}), let us examine that of a
{\it field induced two-angle solution} in a $1D$ chain (cyclic or open) of
spin $s$ with first neighbor $XY$ couplings ($v_\mu^{ij}=\delta_{i,j\pm 1}
v_\mu$, with  $v_z=0$). We assume $\chi=v_y/v_x\in[0,1]$. A separable
eigenstate with $\theta_{2i}=\theta_e$, $\theta_{2i-1}=\theta_o$ is feasible if
there is an alternating field $b^{2i}=b_e$, $b^{2i-1}=b_o$ in inner sites
satisfying (Eqs.\ (\ref{a})--(\ref{b}))
\begin{equation} b_e b_o=(2s)^2v_xv_y\label{f}\,.\end{equation}
This leads to a transverse {\it separability curve}. The ensuing angles satisfy
$\cos\theta_o\cos\theta_e=v_y/v_x$ and are given by
 \begin{equation}
 \cos^2\theta_\sigma=\frac{b_\sigma^2+(2sv_y)^2}
 {b_\sigma^2+(2sv_x)^2}\,,\;\;\sigma=o,e\,,
 \end{equation}
being {\it field dependent}. For $b_e=b_o$ we recover the previous uniform
solution ($b_s=2s\sqrt{v_xv_y}$).
In an open chain we should just add, according to Eq.\ (\ref{b}), the
border corrections $b^1=\half b_o$, $b^n=\half b_{\sigma_n}$. The states
$|\pm\Theta\rangle$ will then be GS setting $\theta_{o,e}>0$ when $v_x>0$ and
$\theta_o>0$, $\theta_e<0$ in the antiferromagnetic case $v_x<0$ (for even $n$
if chain is cyclic, to avoid frustration).

The definite parity states $|\Theta^\pm\rangle$ will again lead to infinite
entanglement range, but with three different field dependent (and hence {\it
controllable}) pairwise concurrences between any two spins (Eq.\ (\ref{Cbc})
for $B=i$, $C=j$): even-even, odd-odd and even-odd, satisfying
$C^\pm_{oe}=\sqrt{C^\pm_{oo}C^\pm_{ee}}$, with $C^\pm_{oo}> C^\pm_{oe}>
C^\pm_{ee}$ if $|b_o|<|b_e|$. Hence, $C^\pm_{oo}$ {\it can be made larger than}
$C^\pm_{oe}$ despite the absence of odd-odd direct coupling. For sufficiently
large $b_e$,  $\cos\theta_e\approx 1$ but $\cos\theta_o\approx \chi$: just
odd-odd pairs will be appreciably entangled in this limit at the SP.

\begin{figure}[t]
\vspace*{0cm}

\centerline{\hspace*{0.cm}\scalebox{.7}{\includegraphics{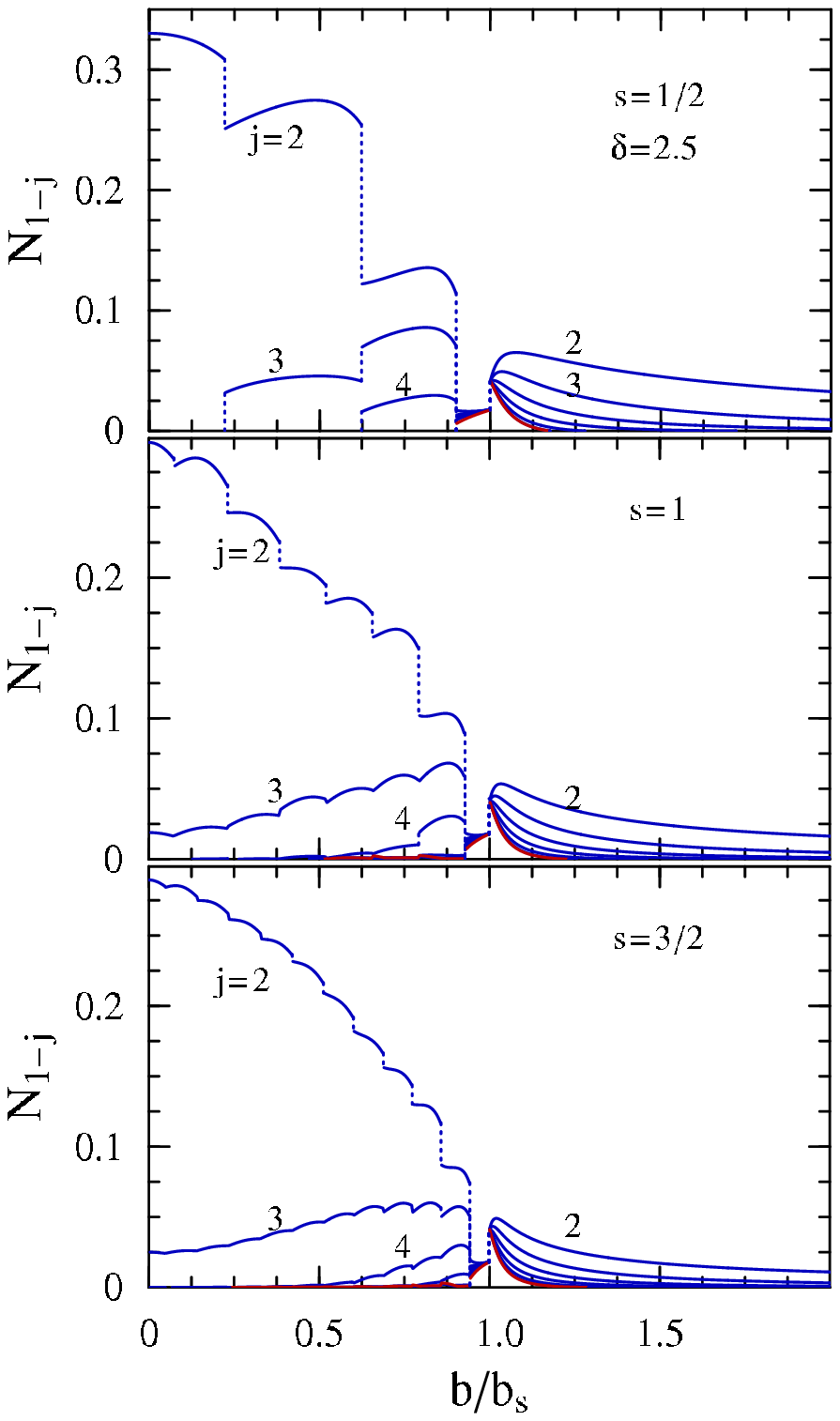}}}
 \vspace*{0cm}

\caption{(Color online) Negativities between the first and the $j^{\rm th}$
spin in an open spin $s$ chain with first neighbor $XY$ coupling, as a function
of the transverse field $b$, with  border corrections (see text), and
three different values of $s$.  We have set an anisotropy
$v_y/v_x=1-\delta/(2sn)$, with $\delta=2.5$ and $n=8$ spins. The factorizing
field corresponds to the last parity transition, and is singled out as the
field where all negativities merge, approaching common non-zero distinct side
limits. The lowest line (in red) depicts the end-to-end negativity
($N_{1-n}$).} \label{f1}
\end{figure}

{\it Application}. As illustration, we first depict in fig.\ \ref{f1} full
exact results for the GS negativities $N_{1j}$ between spins $1$ and $j$ in a
small open chain of uniform spin $s$ with first neighbor $XY$ couplings in a
uniform transverse field $b^i=b$ for $i=2,\ldots,n-1$, with the border
corrections $b^1=b^n=\half b$. For $\chi=v_y/v_x\in(0,1)$ this chain will
then exhibit an exact factorizing field $b_s=2sv_x\sqrt{\chi}$ where separable
parity breaking states with uniform angle $\cos\theta=\sqrt{\chi}$ will become
exact GS if $v_x>0$ (if $v_x<0$, $\theta_i=(-1)^i\theta $ instead in the GS).
We have set $\chi=1-\delta/(2ns)$, such that the side limits of the negativity
at $b_s$ are roughly independent of $s$ and $n$.  It is first seen that the
ensuing behavior of the $N_{1j}$ in terms of the scaled field $b/b_s$ is quite
similar for the three spin values considered ($s=1/2$, $1$ and $3/2$, the
latter involving a diagonalization in a basis of $65536$ states for $n=8$). The
GS exhibits $ns$ parity transitions as the field is increased from $0^+$ to
$b_s$, with the last transition at $b_s$. As the latter is approached, it is
verified that the pairwise entanglement range increases, with {\it all}
negativities approaching the common side limits (\ref{Neg}), distinct at each
side, given here by $N_{ij}^+=\half C_{ij}^+\approx \frac{\delta
e^{-\delta/2}}{2n(1+e^{-\delta/2})}$ (Eq.\ (\ref{cgbcu2})) and $N_{ij}^-\approx
\frac{(C^-_{ij})^2e^{\delta/2}}{4p^-_{A^-}}\approx \frac{\delta^2
e^{-\delta/2}}{4n^2(1-e^{-\delta/2})^2}$. An interval of full range pairwise
entanglement around $b_s$ is then originated, which involves on the left side
essentially  the last state before the last transition (roughly an $W$-state).
$b_s$ plays in this small chain the role of a quantum critical field.

\begin{figure}[t]
\vspace*{0.cm}

\centerline{\hspace*{-1.cm}\scalebox{.55}{\includegraphics{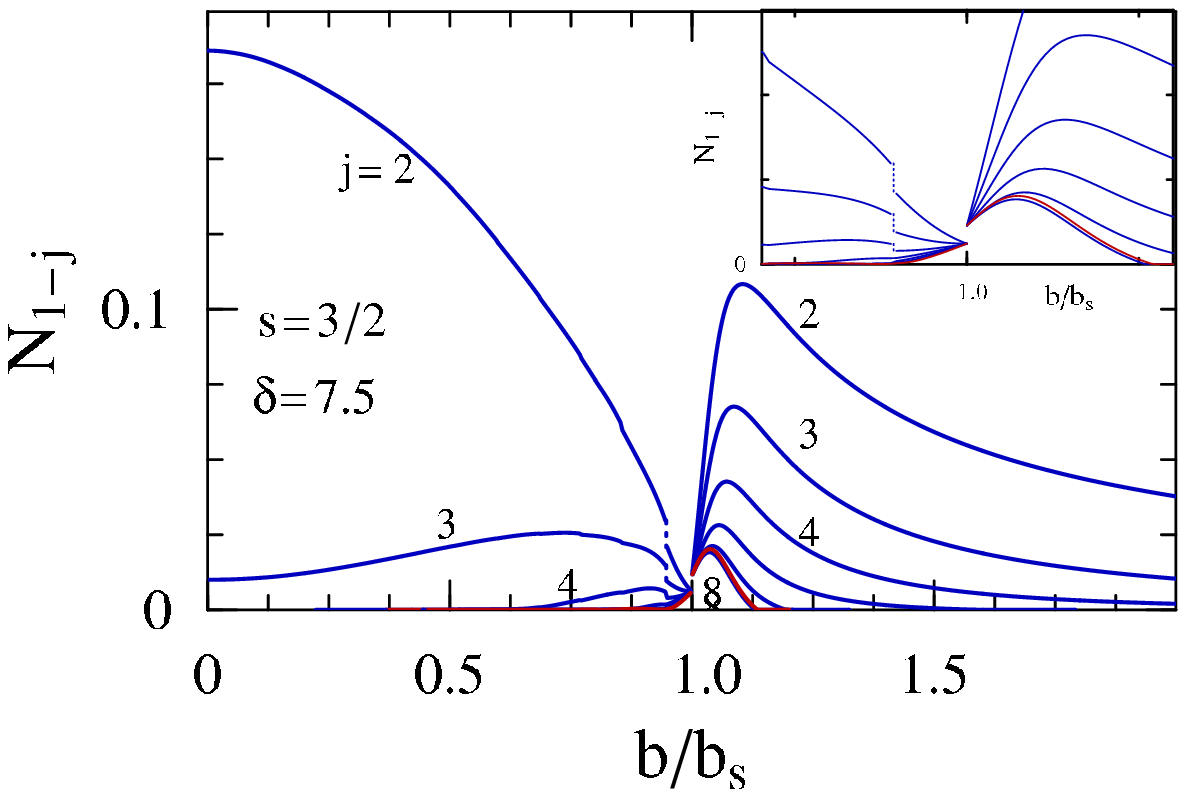}}}
 \vspace*{0.cm}

\caption{(Color online) Same details as fig.\ 1 for $\delta=7.5$ and $s=3/2$.
The inset depicts the behavior in the vicinity of the separability field. }
\label{f2}
\end{figure}
Fig.\ \ref{f2} depicts results for a greater anisotropy $\delta=7.5$. In this
case just the last two parity transitions are  visible in the negativity. The
common side limits of $N^\pm_{ij}$ are smaller and all negativities exhibit a
maximum to the right of the factorizing field. The behavior is then more
similar to that of larger $XY$ systems \cite{AA.06}. Nonetheless, there is
still a clear interval of full entanglement range around $b_s$, with finite
side-limits at $b_s$ when observed in detail (inset).

\begin{figure}[t]
\vspace*{0.cm}

\centerline{\hspace*{-1.cm}\scalebox{.55}{\includegraphics{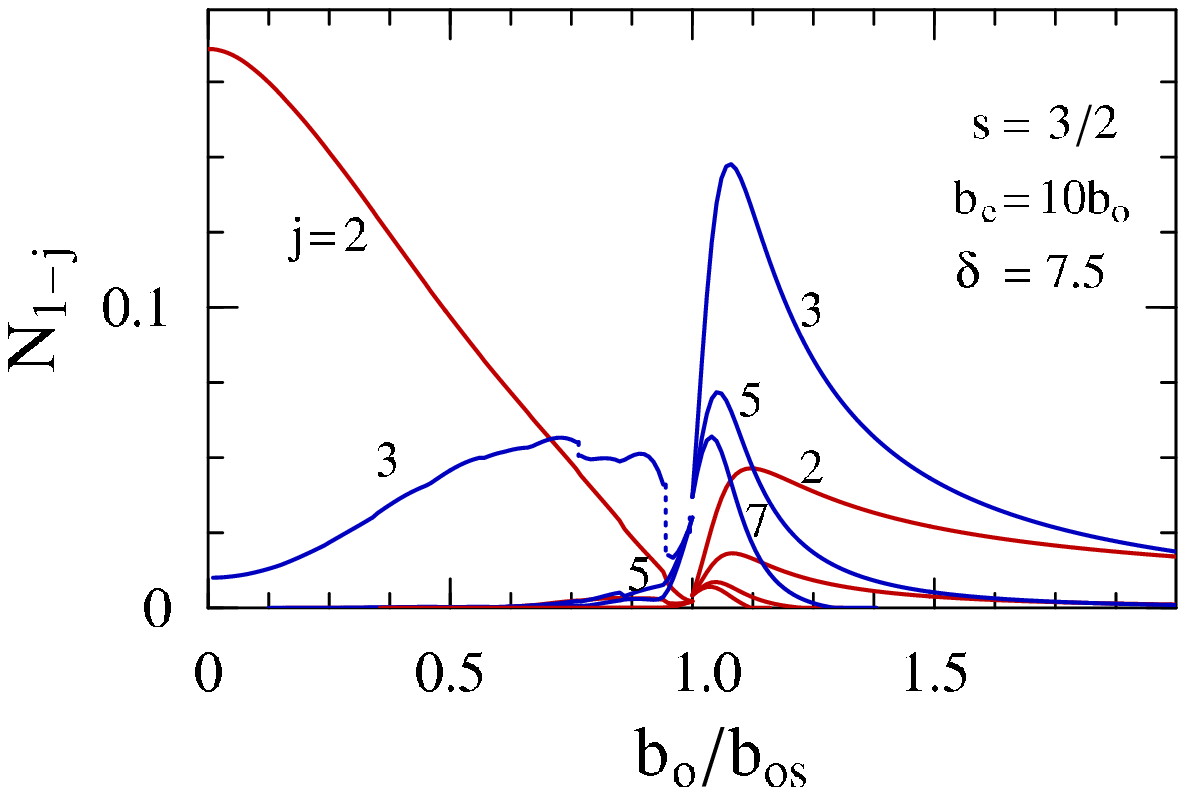}}}
 \vspace*{0.cm}

\caption{(Color online) Same details as fig.\ 1 for $\delta=7.5$, $s=3/2$ and
an alternating field with fixed even/odd ratio $b_e/b_o=10$. Red (blue) lines
depict results for $j$ even (odd).} \label{f3}
\end{figure}

The side limits at separability can actually be modified in this system by
changing the even-odd field ratio $\eta=b_e/b_o$, according to Eq.\
(\ref{f}).  Results for a fixed ratio $\eta=10$ (with pertinent border
corrections) are shown in Fig.\ \ref{f3}, in which case separability is exactly
attained at an odd field $b_{os}=b_s/\sqrt{\eta}$. We have again plotted just
the negativities between the first and the $j$ spin, which now approach {\it
two} common side limits at each side, one for $j$ even ($N^\pm_{oe}$) and one
for $j$ odd ($N^{\pm}_{oo}$). While the former become quite small, the latter
become clearly appreciable, the final effect for such large ratios being
essentially that just odd sites become uniformly entangled in the vicinity of
$b_{os}$. Even-even negativities $N^\pm_{ee}$ (not shown) are of course also
very small at $b_{os}$. Notice finally that $N_{13}$ can become much larger
than $N_{12}$ in the region around  $b_{os}$, despite the absence of second
neighbor couplings.

In summary, we have first determined the conditions for the existence of
separable parity breaking (and locally coherent) eigenstates in general $XYZ$
arrays of arbitrary spins in a general transverse field, showing in particular
the possibility of exact separability in open as well as non-uniform chains
through non-uniform transverse fields. We have also determined the entanglement
properties of the associated definite parity states, through the evaluation of
the concurrence and negativity for any pair of spins or subsystems, for any
spin values. These states, which approach both GHZ and $W$-states in particular
limits, exhibit full entanglement range when non-orthogonal, and can be seen as
effective two qubit entangled states for any bipartition. Moreover, the same
holds for their uniform mixture as well as for the reduced density of any
subsystem. The finite entanglement limits at the SP become relevant in finite
arrays close to the $XXZ$ limit, where the separability field can be clearly
identified with the last GS parity transition, as verified in the numerical
results presented, playing  the role of a quantum critical field. The
possibility of exact separability in an alternating field ($b_e=\eta b_o$) for
arbitrary even-odd ratios $\eta$, leading to controllable entanglement
side-limits, has also been disclosed. The present results provide a deeper
understanding of the behavior of pairwise entanglement in finite XYZ spin
arrays subject to transverse fields.

The authors acknowledge support from CIC (RR) and CONICET (NC, JMM) of
Argentina.


\begin{thebibliography}{999}
\bibitem{r1} In terms of Euler angles,
$e^{i\alpha s^z}e^{i\beta s^y}e^{i\gamma s^z}|0\rangle=ce^{i\theta
s^y}|0\rangle$, with $\tan\frac{\theta}{2}=e^{i\alpha}\tan\frac{\beta}{2}$ and
$c=e^{-is(\alpha+\gamma)}(\frac{\cos\beta/2}{\cos\theta/2})^{2s}$ (Eq.\
\ref{Thex}).  $\bm{\phi}$ can obviously be restricted to the $x,y$ plane.
\bibitem{r2}
For a full isotropic coupling $\half\sum_{i,j}v^{ij}\bm{s}_i\cdot\bm{s}_j$,
$\theta$ remains arbitrary while Eq.\ (\ref{b}) leads to $b^i=0$ if
$\sin\theta\neq 0$: At zero field any global coherent state is here an exact
eigenstate.
\end{thebibliography}
\end{document}